\documentclass[preprint,preprintnumbers,amsmath,amssymb,floatfix]{revtex4}

\usepackage[pdftex]{graphicx}
\usepackage{dcolumn}
\usepackage{bm}

\begin{document}

\title{Experimental Studies of the Jamming Behaviour of Triblock Copolymer Solutions and Triblock Copolymer-Anionic Surfactant Mixtures}

\author{Rajib Basak}
\email{rajib@rri.res.in}
\affiliation{Raman Research Institute, Bangalore 560080, INDIA}
\author{Nabaneeta Mukhopadhyay}
\affiliation{Raman Research Institute, Bangalore 560080, INDIA}
\author{Ranjini Bandyopadhyay}
\email{ranjini@rri.res.in}
\affiliation{Raman Research Institute, Bangalore 560080, INDIA}
\vspace{0.5cm}

\date{\today}

\begin{abstract}
Photon correlation spectroscopy and rheological measurements are performed to investigate the microscopic dynamics and mechanical responses of aqueous solutions of triblock copolymers and aqueous mixtures of triblock copolymers and anionic surfactants. Increasing the concentration of  triblock copolymers results in a sharp increase in the magnitude of the complex moduli characterising the samples. This is understood in terms of the changes in the aggregation and packing behaviours of the copolymers and the constraints imposed upon their dynamics due to increased close packing. The addition of suitable quantities of an anionic surfactant to a strongly elastic copolymer solution results in a decrease in the complex moduli of the samples by several decades. It is argued that the shape anisotropy and size polydispersity of the micelles comprising mixtures cause dramatic changes in the packing behaviour, resulting in sample unjamming and the observed decrease in complex moduli. Finally, a phase diagram is constructed in the temperature-surfactant concentration plane to summarise the jamming-unjamming behaviour of aggregates constituting triblock copolymer-anionic surfactant mixtures. 

\end{abstract}
\maketitle     
\section{Introduction}

When diblock copolymers A$_{x}$B$_{y}$ and triblock copolymers A$_{x}$B$_{y}$A$_{x}$ are dissolved in certain selective solvents, they self-assemble to form micellar aggregates whose properties often mimic those of micelles formed by low molecular weight surfactants \cite{loh_ency}. However, in contrast to normal surfactants, the aggregation of block copolymers depends more strongly on temperature. For triblock copolymers of the type EO$_{x}$PO$_{y}$EO$_{x}$ (EO: ethylene oxide and PO: propylene oxide) that are typically identified by the generic name Pluronic, the hydrophobicity of the central polypropylene oxide (PPO) increases with temperature, while the polyethylene oxide (PEO) blocks at the ends remain hydrophilic \cite{israel_pnas}. Aqueous solutions of these compounds can therefore show significant surface-activity above certain temperatures and concentrations, and can aggregate to form micelles with a dense PPO core surrounded by a corona of hydrated PEO brush-like chains \cite{wan_macro}. Detailed experiments have been performed to study such aggregation behaviour, its dependence on temperature and copolymer concentration, and its effects on the sample dynamics and phase behaviour, using techniques such as small angle x-ray scattering  \cite{caste_jcp,cast_sm}, static  light scattering \cite{lobry_pre,wan_macro}, photon correlation spectroscopy \cite{alex_lang,nystrom_lang,jebari_pol}, small angle neutron scattering (SANS) \cite{cast_sm,lobry_pre,yardim_jcp,mort_macro}, differential scanning calorimetry \cite{wan_macro,alex_lang} and rheology \cite{lobry_pre,prud_lang,caste_jcp,lau_jpol,harsha_pre}.   

 A wide variety of systems, including polymer solutions, granular media, colloidal suspensions and molecular systems, exhibit the jamming phenomenon, characterised by a sudden arrest of their dynamics and the emergence of an elastic response \cite{liu_nat,trappe_nat,corwin_nat}. According to the universal jamming phase diagram proposed by Liu {\it et al.} \cite{liu_nat}, an unjammed system can undergo a transition to a jammed state when its concentration is increased. At high volume fractions, triblock copolymer micelles can enter a solid-like phase that is characterised by very large values of the elastic modulus and very slow dynamics, features that are typically associated with soft glasses \cite{harsha_pre}. Solid-like mesophases of block copolymers exhibit low yield stresses, very high zero shear viscosities ($\sim$ 10$^{6}$ Pa.s) and shear-thinning. Lobry {\it et al.} describe the observed behaviour as a gelation process involving a percolation transition between a micellar liquid phase and a solid phase \cite{lobry_pre}. In contrast, Castelletto {\it et al.} attribute the solidity of the sample to the coexistence of a close-packed crystal with a fluid \cite{caste_jcp}. It is now well-known that in the case of Pluronics of varying chain lengths, the sequence of phase transitions normally remains the same with increasing concentration, while the relative extents of the phases depend upon the details of the copolymer architecture \cite{yardim_jcp}. 

Block copolymers find  important uses in diverse industrial and technical applications and as novel agents in drug and gene delivery \cite{kaba_control}. In this paper, rheological measurements are combined with photon correlation spectroscopy (PCS) experiments to relate the mechanical response of copolymer samples to the microscopic dynamics of the constituent aggregates. With the advent of modern commercial rheometers \cite{macosko}, there have been extensive theoretical and experimental studies to understand the soft glassy rheology of materials as diverse as foams, pastes, emulsions and colloidal suspensions \cite{sollich_prl,sollich_pre,kyu_jnnfn,miyazaki_epl,wyss_prl}. Above a certain copolymer concentration, pure F127 solutions have very high viscosities ($\sim$ 10$^{5}$-10$^{6}$ Pa.s) and can be characterised as soft glasses. In a recent publication, our group established the presence of soft glassy rheology in Pluronic F108 (EO$_{127}$PO$_{48}$EO$_{127}$) solutions, estimated the characteristic relaxation times of the samples and isolated a coexistence of glassy and fluid-like regimes in a temperature-copolymer concentration phase diagram \cite{harsha_pre}. One of the aims of this paper is to show that the emergence of soft glassy rheology in triblock copolymer solutions is accompanied by a slowing down of the slow relaxation mode of the sample, similar to the results reported in \cite{megen_pra,pusey_phya} for concentrated, nonaqueous suspensions of sterically stabilised colloidal spheres. 

Recent experiments have studied the aggregation of Pluronic micelles in the presence of the drug Ibuprofen \cite{foster_lang}. An understanding of the interaction between polymers and surfactants is important due to the wide use of polymer-surfactant mixtures in the manufacture of paints, detergents, cosmetics and pharmaceuticals \cite{jansson_jpc,sastry_cs}. The association of additives with Pluronic molecules is therefore a very interesting topic and has been studied in detail in the literature \cite{hecht_lang,desai_cs,ivanova_cs,li_lang}. Recent studies find that when the anionic surfactant SDS (sodium dodecyl sulfate) is added to Pluronic solutions, the micellar gel structure is altered considerably, and the structural changes observed are strongly dependent on the length ratios of the PO and EO blocks of the Pluronic sample, the concentrations of Pluronic and SDS and the sample temperature. Hecht {\it et al.} show that for F127 (EO$_{70}$PO$_{100}$EO$_{70}$) solutions, the addition of SDS can completely suppress the micellization of Pluronic by destroying the gel phase formed in pure Pluronic micellar solutions \cite{hecht_jpc}, while for P123 (EO$_{21}$PO$_{67}$EO$_{21}$) solutions, the addition of SDS initially enhances the stability of the gel phase before completely destroying this phase at very high SDS concentrations \cite{ganguly_jpc}. It is now well-established that the addition of SDS causes the hydrophobic tail (DS$^{-}$) of the surfactant to bind to the hydrophobic core of the Pluronic micelles \cite{sastry_cs}. The resulting intra-aggregate repulsion causes the breakup of the aggregates into smaller mixed micelles. The formation of such  Pluronic-SDS mixed micelles has been studied in detail using calorimetry \cite{li_lang0,li_lang,thurn_lang}, light scattering \cite{li_lang0,li_lang,mata_cs,jansson_jpc}, electromotive force measurements \cite{li_lang,thurn_lang} and rheology \cite{mata_cs,kurumada_prog}. To investigate the effects of an anionic surfactant additive (SDS) on the micellar packing in F127 solutions, the rheological properties of F127-SDS mixtures are studied by varying the concentration of SDS over a very wide range (mole ratio [SDS]/[F127] is varied between 0 and 35). It is argued here that the shape anisotropy and the size polydispersity of the constituents of F127-SDS mixtures give rise to dramatic changes in the packing behaviour of the micelles, resulting in micellar unjamming and our observation of the disappearance of soft glassy rheology in these samples. This unjamming process causes the complex moduli of the sample to decrease by almost six orders of magnitude in the SDS concentration range studied. This is consistent with our PCS data which shows a speeding up of the relaxation dynamics in F127-SDS mixtures.

\section{Sample preparation and Experimental methods} 

\subsection{Sample Preparation }

Pluronic F127 and SDS (molecular formula: C$_{12}$H$_{25}$SO$_{4}$Na) were purchased from Sigma-Aldrich and used as received without further purification. Pluronic F127 has a molecular weight of 12,600 g/mol, a critical micellization concentration of 0.007 g/cc and a critical micellization temperature between 15-20$^{\circ}$C. SDS, an anionic surfactant with a molecular weight 288.38 g/mol and a critical micellization concentration of 8.2 mM, is selected as previous experiments suggest that SDS binds strongly with F127 \cite{hecht_lang}. To prepare pure F127 solutions, appropriate amounts of F127 are dissolved in deionised and distilled Millipore water (measured resistivity: 18.2 M$\Omega$-cm) under vigorous stirring conditions. To prepare F127-SDS mixtures, appropriate amounts of SDS are dissolved in deionised and distilled Millipore water. Next, F127 is slowly added to the solution and the mixture is stirred in a magnetic stirrer. Each sample is homogenised by storing it overnight at 5$^{\circ}$C, which is well below the critical micellization temperature of F127. The concentration of F127 in all the F127-SDS mixtures is kept constant at 0.25 g/cc (19.8  mM). 
\subsection{ PCS (Photon Correlation Spectroscopy) Measurements}
A BIC (Brookhaven Instruments Corporation) BI-200SM spectrometer is used to measure the light intensity scattered by the samples at angles between $60^{\circ}$ and $135^{\circ}$.  A 150 mW solid state laser (Spectra Physics Excelsior) with an emission wavelength of 532 nm is used as a light source. The sample cell is held in a brass thermostat block filled with decalin, a refractive index matching liquid. The temperature of the sample cell is controlled between 10$^{\circ}$C and 80$^{\circ}$C with the help of a temperature controller (Polyscience Digital). A Brookhaven BI-9000AT Digital Autocorrelator is used to measure the intensity autocorrelation functions of the light scattered from the samples. The intensity autocorrelation function G$^{(2)}(\tau)$ is defined as G$^{(2)}(\tau) = \frac{<I(0)I(\tau)>}{<I(0)>^{2}} =  1+ A|g^{(1)}(\tau)|^{2}$, \cite{bern_pecora},
where $I(\tau)$ is the intensity recorded at time $\tau$, g$^{(1)}(\tau)$ is the normalised electric field autocorrelation function, A is the coherence  factor, and the angular brackets $< >$ represents an average over time. For a dilute solution of monodisperse scatterers,  g$^{(1)}(\tau) \sim \exp(-\tau/\tau_{R})$, where $\tau_R$ is the relaxation time of the scatterers. For a suspension of spheres diffusing in a solvent of refractive index $n$, $\frac{1}{\tau_{R}} = Dq^{2}$, where $D$ is the translational diffusion coefficient and $q={\frac{4\pi{n}}{\lambda}}\sin (\theta/2)$ is the scattering wave vector at a scattering angle $\theta$ and for a wavelength $\lambda$. The effective hydrodynamic radius $R_{H}$ of the scatterer is estimated by using the Stokes-Einstein relation $ D = k_{B}T/6\pi\eta{R_{H}}$, where $k_{B}$ is the Boltzmann constant, $T$ is the temperature and $\eta$ is the viscosity of the solvent. For a suspension of polydisperse particles, if the diffusive relaxation times $\tau_{R}$ are closely spaced, then  g$^{(1)}(\tau) \sim \exp(-\tau/\tau_{R})^{\beta}$, where $\beta <$ 1.  
\subsection{ Rheological Measurements}
All the rheological measurements are performed in a modular compact rheometer Anton Paar MCR 501. These experiments are performed in a double gap geometry DG26.7/Q1 which can measure shear stresses in the range between 109.407$\times$10$^{-6}$ Pa and 2516.361 Pa, shear strains above 2.93$\times$10$^{-5}$ and strain rates between 3.0697$\times$10$^{-7}$s$^{-1}$ and 9.209$\times$10$^{3}$s$^{-1}$. This geometry has a gap of 1.886 mm, an effective length of 40 mm and requires a sample volume of 3.8 ml for every run. A water circulation unit Viscotherm VT2 is used to control the temperature of the sample in the measuring cell in the range between 5$^{\circ}$C and 80$^{\circ}$C. For each run, a fresh sample was loaded and  rheological data was acquired using the Rheoplus software (version 3.40) provided by the manufacturer.

\section{Results} 
\begin{figure}
\begin{center}
\includegraphics[width=4in]{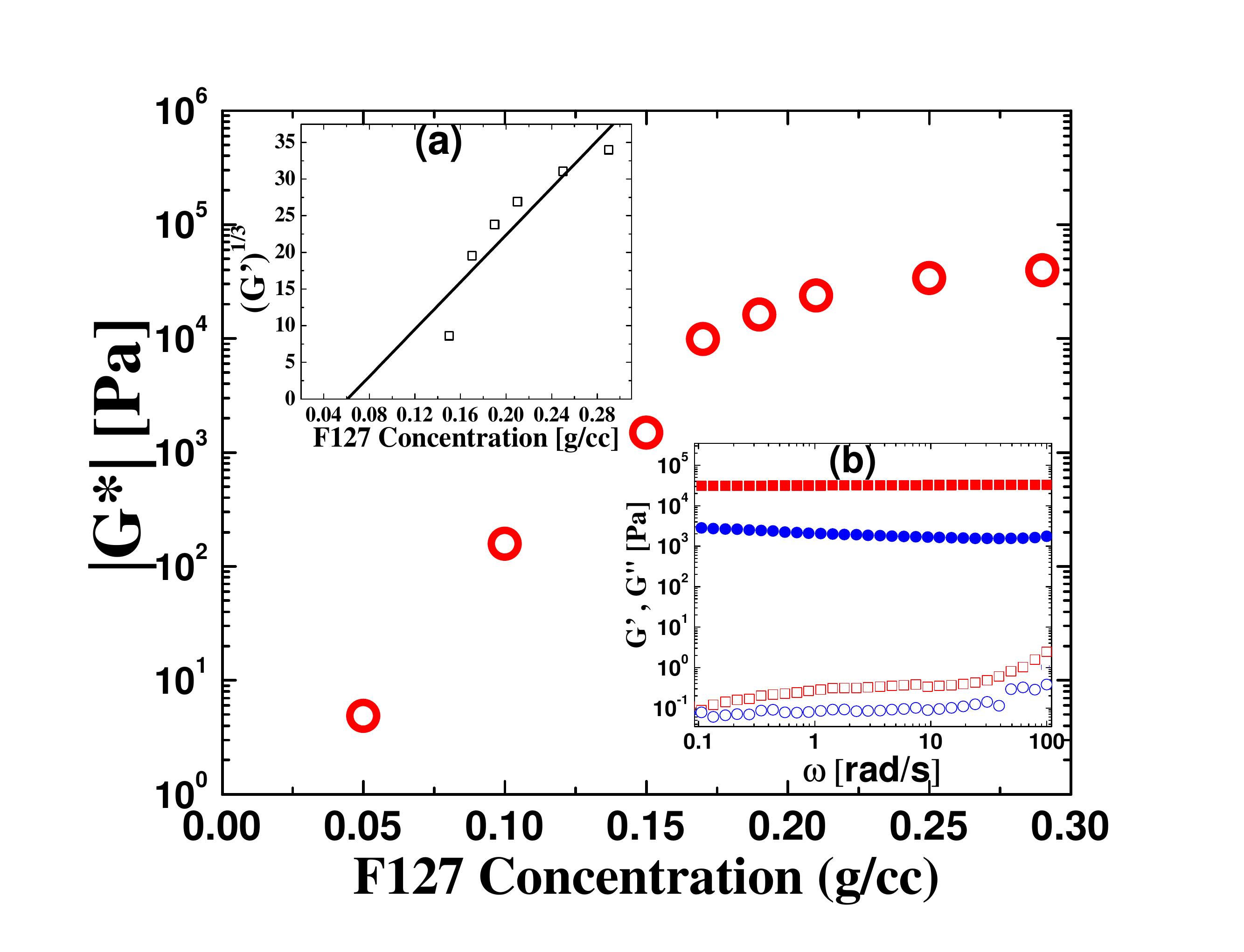}
\caption{The magnitude of the complex modulus $|{G^{\star}}|$ ($\circ$) at T = 40$^{\circ}$C with increasing F127 concentration. Inset (a) shows the plot of ${G^{\prime}}^{1/3}$ {\it vs.} F127 concentration. A straight line fit to the data intersects the {\it x} axis at $c_{gel}$ = 0.06 g/cc. Inset (b) shows the frequency responses at 40$^{\circ}$C for F127 solutions of concentration (a) 0.05 g/cc (hollow symbols) and (b) 0.25 g/cc (solid symbols). The  squares denote G$^{\prime}$ and the  circles denote G$^{\prime\prime}$.} 
\label{Fig. 1}
\end{center}
\end{figure}
Fig. 1  plots the magnitude of the complex modulus $|G^{\star}|$ (circles) of F127 solutions {\it vs.} F127 concentration.  $|G^{\star}|$ is defined as $|G^{\star}| = {(G^{{\prime}2} + G^{{\prime\prime}2})}^{1/2}$ \cite{macosko}, where G$^{\prime}$ and G$^{\prime\prime}$ are the elastic (storage) and viscous (loss) moduli respectively at an angular frequency $\omega$ = 1 rad/s, acquired in oscillatory rheology experiments performed at small strain amplitudes $\gamma =$ 0.5\% at 40$^{\circ}$C. It can be seen from Fig. 1 that $|G^{\star}|$ shows a monotonic increase with F127 concentration, similar to the results in \cite{wanka_polsci}. At the lowest concentration of F127 (0.05 g/cc), $|G^{\star}|$ $\approx 5$ Pa.  When the concentration of F127 is increased to 0.2 g/cc, $|G^{\star}|$ increases by 4 decades ($\approx 5 \times 10^{4}$ Pa).  Similar to the work on near critical polymer gels reported in \cite{colby_pre}, a straight line fit to a plot of ${G^{\prime}}^{1/3}$ {\it vs.} F127 concentration (inset (a) of Fig. 1) intersects the ${G^{\prime}}^{1/3}$ = 0 axis at $c_{gel} \approx$ 0.06 g/cc, which gives a reasonable estimate for the gelation concentration threshhold of F127 micelles. Oscillatory frequency sweep measurements for F127 solutions of different concentrations are performed at T = 40$^{\circ}$C by decreasing the angular frequency $\omega$ logarithmically from 100 rad/s to 0.1 rad/s while keeping the strain amplitude fixed at 0.5\%. For the samples of higher concentrations (data for 0.25 g/cc F127 solution denoted by solid symbols in inset (b) of Fig. 1), the elastic modulus G$^{\prime}$ (denoted by squares) is almost independent of $\omega$, the viscous modulus G$^{\prime\prime}$ (denoted by circles) is weakly dependent on frequency and G$^{\prime}$ $>>$ G$^{\prime\prime}$. These are typical signatures of jammed soft solids \cite{harsha_pre, wyss_prl, miyazaki_epl} and indicate the presence of a crowded micellar environment in this F127 sample. Previous small angle neutron scattering experiments have demonstrated that the aggregation number of F127 micelles are independent of concentration and temperature above the critical micellization temperature \cite{prud_lang}. The increased overlapping of the PEO chains \cite{lobry_pre} in  concentrated F127 solutions therefore results in micellar jamming and compaction and the large magnitudes of the characteristic moduli. In contrast to the high-concentration regime, the magnitudes of G$^{\prime}$ and G$^{\prime\prime}$ (hollow squares and circles respectively in the inset (b) of Fig. 1) are significantly lower in the dilute samples and indicates the absence of a predominantly  solid-like response in this concentration regime. 

\begin{figure}
\begin{center}
\includegraphics[width=6in]{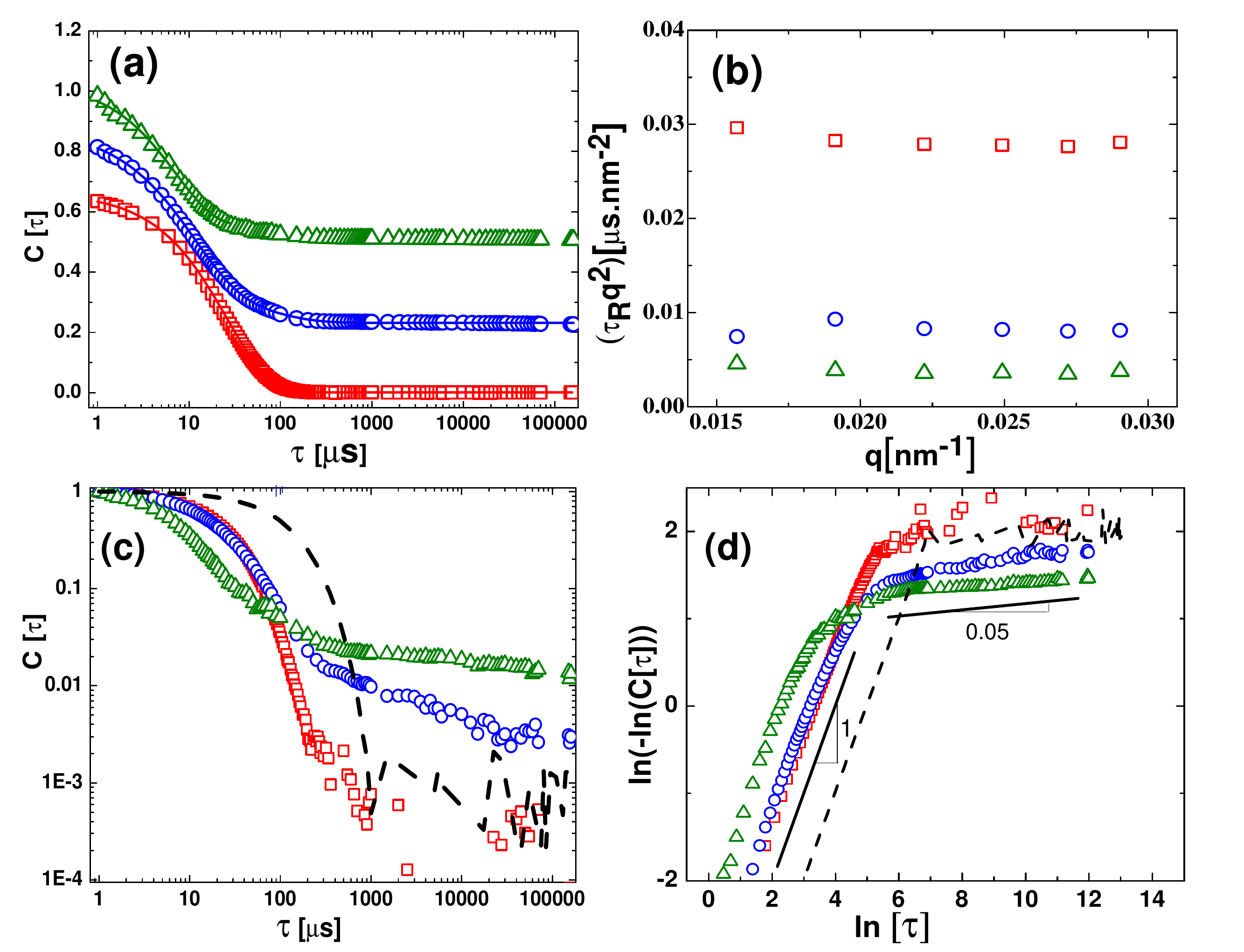}
\caption{ PCS data for F127 solutions of concentrations 0.05 g/cc ($\square$), 0.10 g/cc ($\circ$) and 0.25 g/cc ($\triangle$). 2(a) shows the plots of the  intensity autocorrelation functions C[$\tau$] on a linear-logarithmic scale at T = 40$^{\circ}$C, acquired at $\theta$ =  90$^\circ$ (data sets are shifted vertically for better visibility). Fits to this data, whose functional forms are discussed in the text, are shown by solid lines. Fig. 2(b) shows that  $\tau_{R}q^{2}$  does not change with $q$ for all three samples. 2(c) shows the plots of the normalised intensity autocorrelation functions C[$\tau$] {\it vs.} $\tau$ on a logarithmic-logarithmic scale. The dashed line in (c) corresponds to the autocorrelation plot for a dilute aqueous solution of freely diffusing polystyrene colloidal spheres of size 95 nm at T = 40$^{\circ}$C and $\theta$ =  90$^\circ$. The data of Fig. 2(c) is recast in 2(d), where $\ln(-\ln(C[\tau]))$ is plotted {\it vs.} $\ln[\tau]$. \\}
\label{Fig. 2}
\end{center}
\end{figure}

In order to investigate the relaxation processes in the different concentration regimes of F127 solutions, systematic PCS experiments are performed. To take care of sample non-ergodicity, the method proposed by van Megen and Pusey \cite{pusey_phya} is employed to measure the intensity autocorrelation functions of the light scattered by such samples. The unnormalised time-averaged intensity autocorrelation functions are measured for six different scattering volumes obtained by rotation and vertical translation of the sample cell. The intensity autocorrelation functions acquired thus are added and then normalised. Ensemble averaged autocorrelation functions are acquired at six different scattering angles. For a pure F127 sample of concentration 0.05 g/cc (the low viscosity regime), the intensity autocorrelation function $ C[\tau] = \frac{G^{(2)}(\tau)-1}{A}$ is plotted $vs.$ the delay time $\tau$ at T = 40$^{\circ}$C at scattering angle $\theta$ = 90$^{\circ}$ (squares) in Fig. 2. Fig.  2(a) shows the fit (solid line) of the autocorrelation function $C[\tau]$ to $C[\tau] \sim \exp(-\frac{\tau}{\tau_R})^{\beta}$, where the stretching exponent $0.9 <\beta< 0.95$ indicates the presence of an approximately  single relaxation rate. Fig. 2(b) shows  that  $\tau_{R}q^2$ (squares) does not change with $q$, which confirms the diffusive nature of the micellar relaxation in the dilute regime. 

As the concentration of F127 is increased, the intensity autocorrelation functions do not show stretched exponential decays, but can instead be described as two-step relaxation processes. Fig. 2(a) also shows the intensity autocorrelation plot for a sample in the semi-dilute viscosity regime (0.10 g/cc F127 solution, denoted by circles) at $\theta$ = $90^{0}$ and T = 40$^{\circ}$C. This autocorrelation data is fit to the form $ C[\tau]= (A\exp(-(\tau/\tau_{R})) +B\exp(-(\tau/\tau_{2})^{\beta})^{2}$. The value of $\tau_{R}$ (relaxation time for the fast exponential decay) and $\tau_{2}$ (relaxation time for the slow stretched exponential process) are estimated for six different wavevectors. Fig. 2(b) shows the plot of $\tau_{R}q^{2}$ (denoted by circles) {\it vs.} $q$ for this sample. $\tau_{R}q^{2}$ does not change with $q$, indicating that the fast relaxation of the micelles in the semi-dilute regime is diffusive. The slower stretched exponential relaxation process yields a non-diffusive relaxation time ($\tau_{2}$) that is approximately 10 times slower than $\tau_{R}$. $\tau_{2}$ has an approximately $q^{-3}$ dependence (not shown) and can be attributed to intraparticle dynamics arising out of large scale sample heterogeneities/ clusters \cite{nystrom_lang}. 
 
As the F127 concentration is further increased, the overlapping between the PEO coronas increase and the close-packed micelles undergo a jamming transition due to the constraints imposed upon their motion. The micelles are now trapped in cages formed by their neighbours and structural relaxation by micellar rearrangements becomes less probable. At F127 concentrations $\geq $ 0.15 g/cc, the system enters a 'strongly elastic' soft solid phase characterised by $G^{\prime} >> G^{\prime\prime}$. The fluctuations in the intensity of the light scattered by these samples become so slow that the full decay of C$(\tau)$ to zero (or to instrumental noise level) is not possible within the experimental time window. To highlight the non-ergodic nature of the relaxation at high sample concentrations, the normalised intensity autocorrelation function C[$\tau$] is plotted $vs.$ the delay time $\tau$ on a logarithmic-logarithmic scale in Fig. 2(c). Fig. 2(c) also shows the intensity autocorrelation data for freely diffusing 95 nm polystyrene (PS) spheres in an aqueous solution (volume fraction $\phi$ $\sim$ $10^{-5}$; shown by dashed line). After an initial exponential decay, the C[$\tau$] plot for the freely diffusing PS spheres decays to $10^{-3}$-$10^{-4}$ at $\tau \approx$ 1000 $\mu$s  which defines the instrumental noise level for our experiments.  For the dilute unjammed F127 sample of concentration 0.05 g/cc (squares in Fig. 2(c)), C$(\tau)$ decays to the experimental noise level in $\tau \sim$ 1000$\mu$s, verifying that the dynamics in the dilute concentration regime is ergodic. From the rheological data for the F127 solution of concentration 0.05 g/cc (hollow symbols in inset (b) of Fig. 1), we observe that $G^{\prime}$ $\geq$ $G^{\prime\prime}$, which indicates a weak gel.  However, the PCS data for the same sample (Fig. 2) shows a complete decay of $C(\tau)$, confirming fluid like mobility of the micelles. We believe that this discrepancy arises because rheology and PCS experiments probe very different length scales  \cite{band_SM}.

\noindent For the soft solid-like sample (0.25 g/cc F127, triangles in Fig. 2(c)), it is seen that C[$\tau$] does not decay to the noise level within our experimental time window. This indicates a slowing down of the dynamics of the sample and is a signature of a jammed micellar environment. The data of Fig. 2(c) is recast in Fig. 2(d), where  $\ln(-\ln(C[\tau]))$ is plotted {\it vs.} $\ln[\tau]$ for the same samples. Similar to the PS suspension data (dashed lines), the data for the 0.05 g/cc sample (squares) is a straight line of slope $\beta$ = 1 at $\ln[\tau] \le$ 5 followed by a noisy plateau at $\ln(-\ln(C[\tau])) \sim$ 2, which corresponds to $C(\tau) \approx$ 10$^{-3.5}$ (noise level defined in Fig. 2(c)). The data for the 0.1 g/cc and 0.25 g/cc samples (circles and triangles respectively), are characterised by two components, a faster straight line component which extends upto 2 $\le \ln[\tau] \le$ 4 with a slope $\beta$ = 1. This indicates a mono-exponential faster relaxation process and is followed by a slower straight line component of significantly smaller slope  The systematic decrease in the magnitudes of $\ln(-\ln(C[\tau]))$ and the decreasing slopes of the straight lines characterising the slow components on increasing F127 concentration is a clear indication of dramatic slowing down of the sample dynamics.

In Fig. 2(a), the faster decay of the correlation function for the 0.25 g/cc sample (triangles) is fitted to an exponential form $C[\tau] \sim \exp(\frac{-\tau}{\tau_{R}})$. That the relaxation process associated with this exponential decay is diffusive is verified by the constant values of $\tau_{R}q^{2}$ with changing $q$ (triangles in Fig. 2(b)). In the soft solid phase, the micellar aggregates are closely packed and the interaction between aggregates is very strong \cite{nilsson_jpc}.  The diffusive fast relaxation process is attributed to the constrained diffusion of the confined F127 micelles.  

\begin{figure}
\begin{center}
\includegraphics[width=4in]{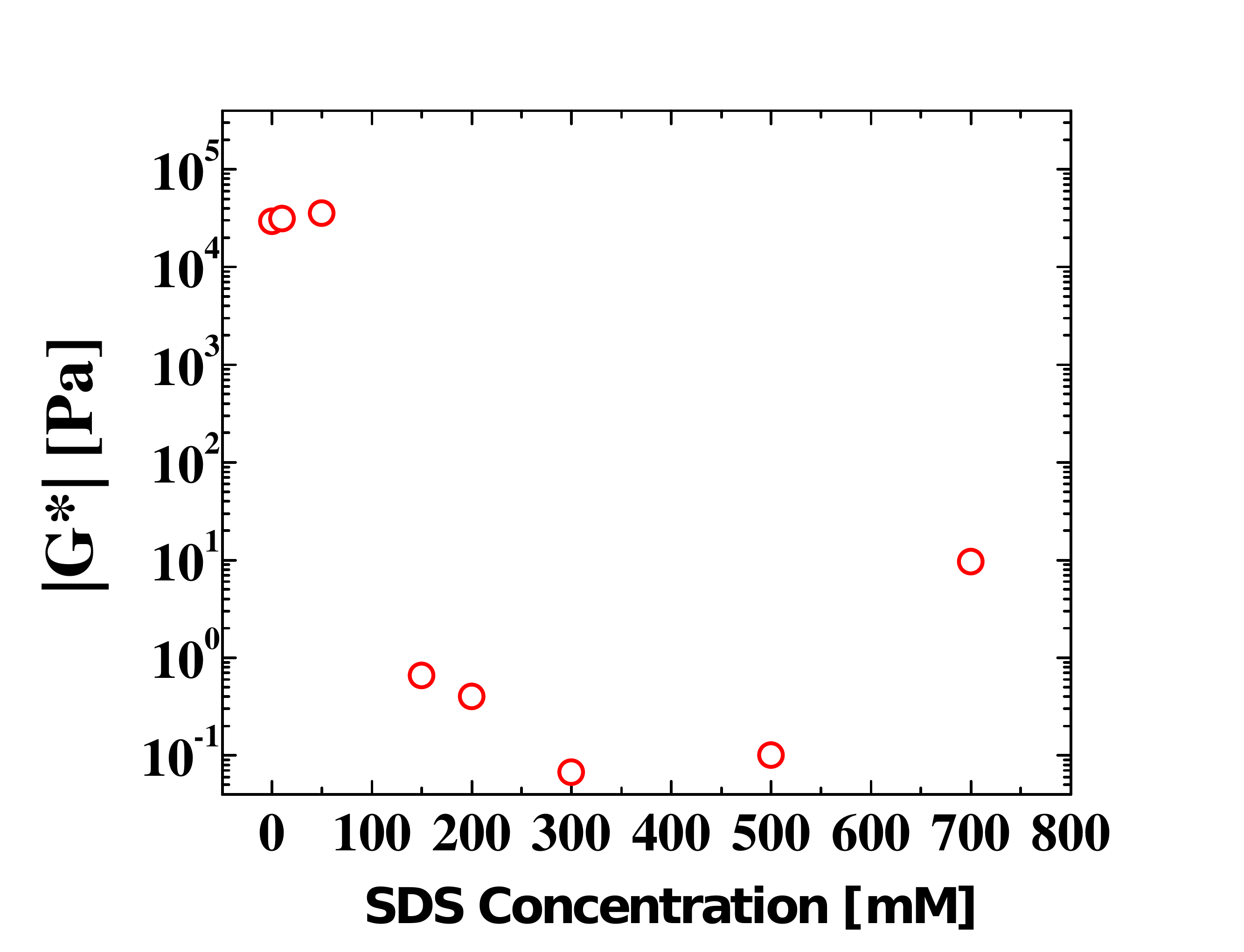}
\caption{The magnitude of the complex modulus $|G^{\star}|$ $(\circ)$  of F127-SDS mixtures (concentration of F127 fixed at 0.25 g/cc) at T = 40$^{\circ}$C with increasing SDS concentration.\\}
\label{Fig. 3}
\end{center}
\end{figure}

Next, F127-SDS mixtures are prepared to study the effects of the addition of an anionic surfactant SDS to the soft solid-like phase formed by concentrated F127 solutions. Fig. 3 plots the magnitudes of the complex moduli $|G^{*}|$ (denoted by circles) of F127-SDS mixtures (concentration of F127 fixed at 0.25 g/cc, SDS concentration varying in the range  0-700 mM) at 40$^{\circ}$C, acquired in oscillatory rheology experiments ($\gamma$ = 0.5\% and $\omega$ = 1 rad/s). As demonstrated earlier, the 0.25 g/cc F127 system is a jammed, soft solid. F127-SDS mixtures exhibit very high magnitudes of the complex moduli $|G^{*}|$ when SDS concentration $\leq$ 50 mM. For higher concentrations of SDS ($\sim$ 300-500 mM), $|G^{*}|$ decreases  by six orders of magnitude. The non-monotonic dependence of the $|G^{\star}|$ on SDS concentration at SDS concentrations $>$ 150 mM arises due to the complex dependence of the mechanical moduli of the mixtures on temperature. 
\begin{figure}
\begin{center}
\includegraphics[width=4in]{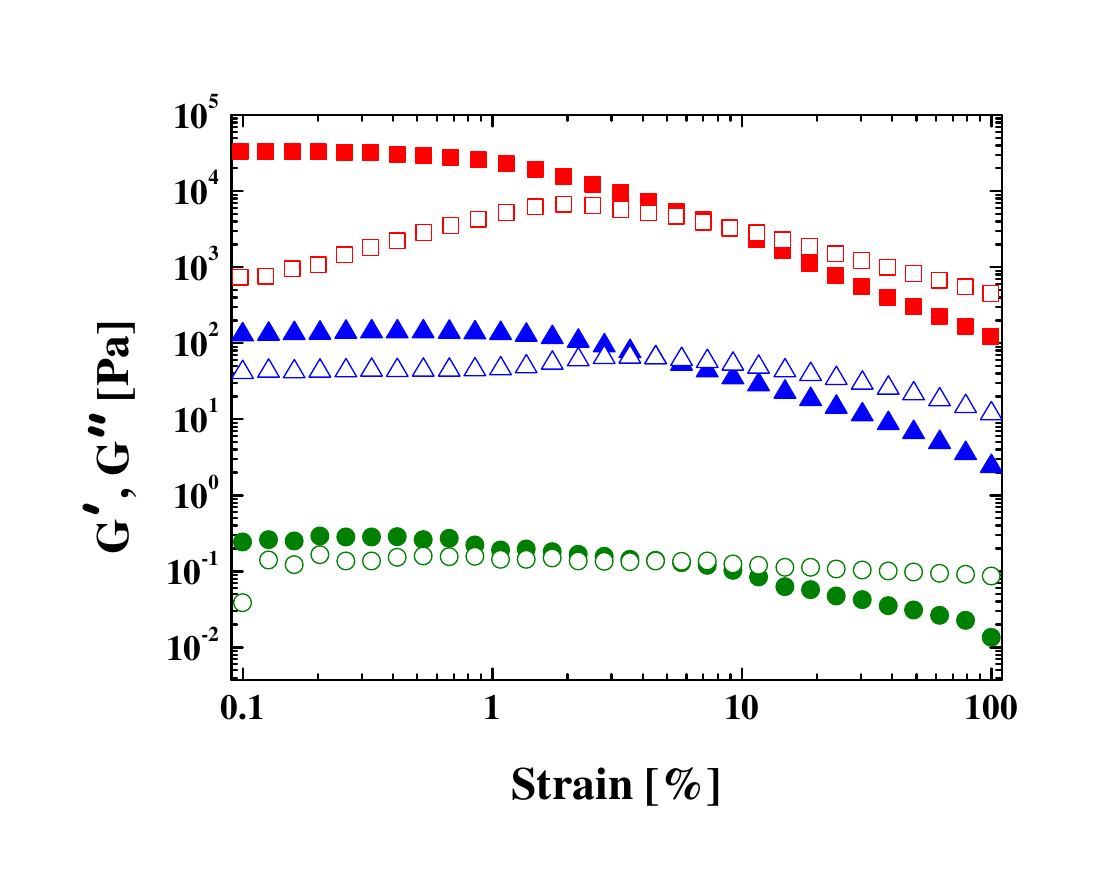}
\caption{Oscillatory amplitude sweep data for  a pure F127 sample ($\square$) and for F127-SDS mixtures  with SDS concentrations 150 mM ($\triangle$) and 700 mM ($\circ$) at T = 40$^{\circ}$C. The F127 concentration is fixed at 0.25 g/cc. The solid symbols denote G$^{\prime}$ and the hollow symbols denote G$^{\prime\prime}$. \\}
\label{Fig. 4}
\end{center}
\end{figure}

 To further characterise the mechanical responses of the mixtures, G$^{\prime}$ and G$^{\prime\prime}$ are measured while performing strain amplitude sweep measurements at T $=$ 40$^{\circ}$C. These experiments are performed by ramping up the amplitude of the oscillatory strain $\gamma$ at a fixed angular frequency $\omega$ = 1 rad/s. The results are displayed in Fig. 4. 
For the pure F127 solution (denoted by squares in Fig. 4), G$^{\prime}$ (filled squares) stays almost constant at the lower strain amplitudes, followed by a power law decrease at strain amplitudes $\gamma > $ 1\%. G$^{\prime\prime}$ (hollow squares) is significantly lower than G$^{\prime}$ and shows a peak at $\gamma \sim$ 2\%, followed by a power law decay . These features, {\it viz.}, G$^{\prime}$ $>>$ G$^{\prime\prime}$ at low strains, a peak in G$^{\prime\prime}$ at a characteristic strain, and power law decays of both moduli at higher strains such that G$^{\prime\prime} >$ G$^{\prime}$ for very high strains, are typical features of soft solids \cite{sollich_prl,kyu_jnnfn,wyss_prl,miyazaki_epl}. With increasing SDS concentration, $G^{\prime}$ and $G^{\prime\prime}$ decrease significantly, the magnitude of G$^{\prime}$ approaches that of G$^{\prime\prime}$ and the peak in G$^{\prime\prime}$ eventually disappears, signalling the  gradual disappearance of soft glassy rheology and the onset of sample unjamming. For the mixture with 700 mM SDS, this is clearly illustrated by the very weak strain dependence of G$^{\prime}$ and G$^{\prime\prime}$ (circles in Fig. 4) and the observation that G$^{\prime}$ $\approx$ G$^{\prime\prime}$ for $\gamma <$ 10\%.

 \begin{figure}
\begin{center}
\includegraphics[width=6in]{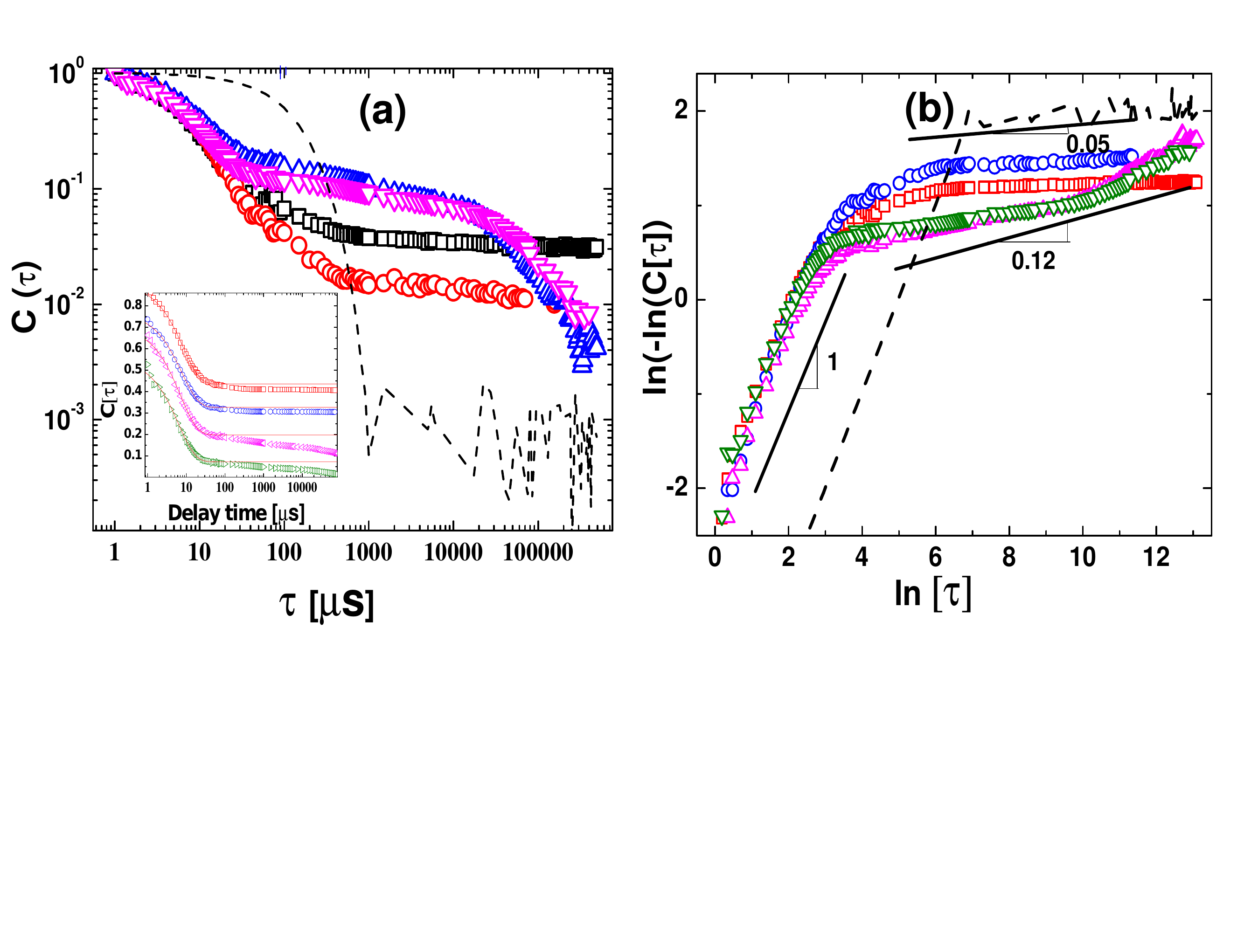}
\caption{In (a), the normalised intensity autocorrelation functions C[$\tau$] are plotted on a logarithmic-logarithmic scale for a pure F127 sample ($\square$), for F127-SDS mixtures containing 10 mM ($\circ$), 300 mM ($\triangle$), 500 mM ($\nabla$) SDS and a dilute aqueous solution of 95 nm polystyrene spheres (shown by dashed line) at T = 40$^{\circ}$C and $\theta$ =  90$^\circ$. The inset shows the linear-logarithmic plots of C[$\tau$] {\it vs.} $\tau$ for the same data sets, with the corresponding exponential fits to the faster relaxation processes shown by solid lines (data sets are shifted vertically for better visibility). The data of 5(a) is replotted as $\ln(-\ln(C[\tau]))$ {\it vs.} $\ln[\tau]$ in 5(b).}
\label{Fig. 5}
\end{center}
\end{figure} 
The microscopic relaxation processes contributing to the unjamming of F127-SDS mixtures are studied by performing PCS experiments. In Fig. 5(a), the normalised intensity autocorrelation function C$(\tau)$ is plotted {\it vs.} the delay time $\tau$ at $\theta$ =  $90^{\circ}$ and T = 40$^{\circ}$C for a pure F127 solution (denoted by squares) and for F127-SDS mixtures with SDS concentrations 10 mM (circles), 300 mM (up triangles) and 500 mM (down triangles) respectively. The PCS plot for freely diffusing 95 nm PS spheres in water is also incorporated as a reference process that exhibits a complete decay to the experimental noise level (shown by dashed line). The $C[\tau]$ plots for the jammed samples (the F127 solution and the F127-SDS mixture with 10 mM SDS, denoted by squares and circles respectively) do not show complete decays to the noise level within the experimental time window. This confirms the slowing down of the slow relaxation times of these samples due to the kinetic constraints experienced by the closed-packed micelles. For F127-SDS mixtures with higher concentrations of SDS (300 mM and 500 mM, denoted by up triangles and down triangles respectively), C$[\tau]$ shows a comparatively faster decay within the experimental time window, indicating that the relaxation times of these mixtures are much faster than those with lower SDS concentrations. The data of Fig. 5(a) is recast in Fig. 5(b) where $\ln(-\ln(C[\tau]))$ is plotted {\it vs.} $\ln[\tau]$. For all the samples, the initial part of the plot comprises a straight line of slope 1, indicating a fast monoexponential component in the sample dynamics. For the pure F127 solution and the F127-SDS mixtures, this plot shows a transition  at $\ln[\tau] \approx$ 3 to a second slower regime which is characterised by a change in the slope of the straight line to very small values.  For the 0.25 g/cc F127 solution (squares) and the F127-SDS mixture with 10mM SDS, the slope of the straight line in the slow regime is approximately 0.05. For the F127-SDS mixtures with 300mM and 500mM SDS  (denoted by up triangles and down triangles respectively), the slow regime is characterised by a more complex functional form, with the data eventually increasing rapidly to the experimental noise level set by the PS spheres at $\ln[\tau] \ge$ 10. This indicates a speeding up of the slow dynamics in these samples.     

It is now understood that the addition of SDS to F127 solutions results in the adsorption of the DS$^{-}$ chains to the hydrophobic PPO cores of the F127 micelles. This imparts negative charges to the mixed aggregates. The resulting intra-aggregate repulsion leads to the breakdown of the large spherical micelles into smaller mixed micelles comprising both F127 and SDS \cite{hecht_lang,sastry_cs}. Previous experiments \cite{jansson_jpc} suggest that the addition of small amounts of SDS can result in the formation of large copolymer-rich charged complexes, while the addition of larger amounts of SDS results in the breakup of these complexes and the formation of smaller surfactant-rich complexes. For high concentrations of SDS, the mixtures comprise anisotropic SDS-rich mixed micelles and pure SDS micelles of smaller sizes \cite{hecht_lang,jansson_jpc}. In contrast to pure copolymer solutions where the micellar aggregates are spherical and monodisperse, aggregates in mixtures are characterised by substantial anisotropy and polydispersity.  Anisotropic particles possess more rotational degrees of freedom than spherical particles and therefore require higher numbers of contacts for mechanically stable close packings. The packing volume fraction required for the random close packing ($\phi_{RCP}$) of anisotropic particles is therefore substantially larger than the $\phi_{RCP}$ of spherical, monodisperse particles \cite{donev_pre,donev_sci}.
Furthermore, simulations of colloidal hard sphere systems \cite{sch_jsm,yang_pre}, theoretical models \cite{ouch_ind} and experimental studies \cite{sohn_can,suz_powder} show that $\phi_{RCP}$ increases with increase in the standard deviation of the particle size. Since smaller spheres can fill the gaps between larger particles very efficiently in packings of polydisperse spheres, the increase of $\phi_{RCP}$ with polydispersity is not surprising. Changes in SDS content in F127-SDS mixtures result in changes in the number fractions of the different micellar species in solution. This induces large changes in the packing behaviours of the aggregates. As a result, F127-SDS mixtures are characterised by $\phi_{RCP}$ values that are higher than those expected for pure F127 solutions. Increase in $\phi_{RCP}$ removes the space constraints previously experienced by the overlapping micelles and results in the unjamming and decompaction of the aggregates in F127-SDS mixtures. This manifests as a speeding up of the microscopic dynamics and the disappearance of soft glassy rheology of the mixtures as SDS content is increased. 

In the inset of Fig. 5(a), the intensity autocorrelation functions are re-plotted on a linear-logarithmic scale (the data is shifted vertically for better visibility) and the fast decays are fitted to the form $C[\tau] \sim \exp(-\frac{\tau}{\tau_{R}})$ for six different wavevectors. It is observed that $\tau_{R}q^{2}$ does not change with $q$ (not shown), confirming the diffusive nature of the fast relaxation processes. For the jammed samples (pure F127 solution and the F127-SDS mixture with 10 mM SDS, denoted by squares and circles in Figs. 5(a) and (b)), the fast processes are attributed to the constrained diffusion of confined micelles. In the F127-SDS mixtures with 300 mM and 500 mM SDS (denoted by up triangles and down triangles in Fig. 5), the constituent aggregates tend to unjam due to the anisotropy and the polydispersity of the constituent aggregates. At high concentrations of SDS, a significant fraction of SDS molecules exist as pure SDS micelles {\cite{hecht_lang} which, due to their small sizes, can diffuse easily through the free spaces available in the system. By combining the Stokes Einstein relation with the measured diffusive relaxation rates for the mixtures with high SDS content, it is estimated that the fast exponential decays are due to the relaxation of scatterers of size $\sim$ 1-2 nm \cite{hecht_lang,almgren_jpc} which can be identified as spherical SDS micelles.  

\begin{figure}
\begin{center}
\includegraphics[width=4in]{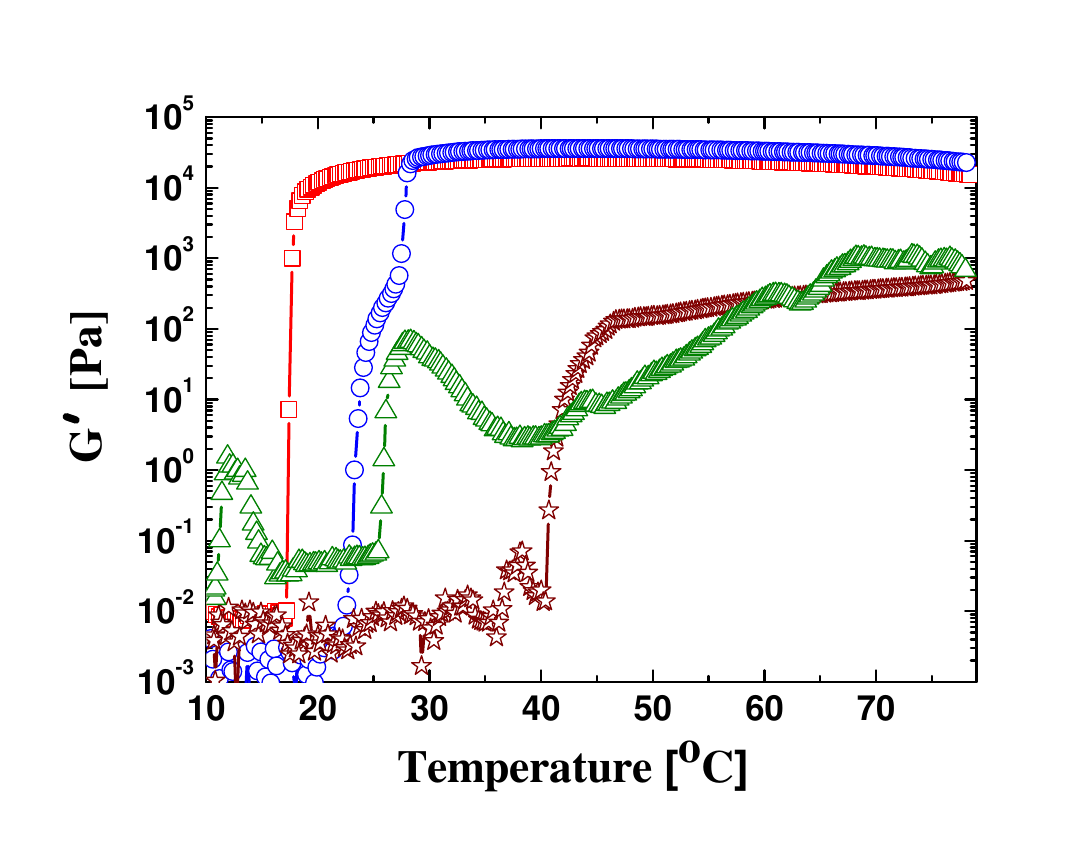}
\caption{Temperature sweep data showing the evolution of G$^{\prime}$ in an F127 solution of concentration 0.25 g/cc ($\square$) and when   50 mM ($\circ$), 150 mM ($\star$) and 700 mM ($\triangle$) SDS is added to the pure F127 solution. \\}
\label{Fig. 8}
\end{center}
\end{figure}

Temperature sweep oscillatory rheology experiments are next performed by ramping up the sample temperatures at a rate of 0.25$^{\circ}$C/minute, while keeping the amplitude of the oscillatory strain constant at $\gamma =$ 0.5\% at an angular frequency $\omega =$ 1 rad/s. This data is plotted in Fig. 6. At the lowest temperatures, the samples comprise unimers in solution and are characterised by low values of the elastic modulus: G$^{\prime} \sim$ 1-10 mPa and G$^{\prime} <$ G$^{\prime\prime}$ (G$^{\prime\prime}$ data is not shown in Fig. 6). Above a certain temperature, each sweep is characterised by an abrupt increase of the elastic modulus G$^{\prime}$ to a higher value that decreases sharply with SDS concentration. 
The increase in G$^{\prime}$ is identified with a liquid-solid phase transition and is consistent with the data reported in our previous publication \cite{harsha_pre}. Our observation that the temperature sweep data of the mixtures are characterised by lower values of G$^{\prime}$ at temperatures greater than the liquid-solid transition temperature is consistent with the picture of micellar unjamming in mixtures proposed earlier. This conclusion is supported by the rheology and PCS data displayed earlier in Figs. 4 and 5. 

\begin{figure}
\begin{center}
\includegraphics[width=4in]{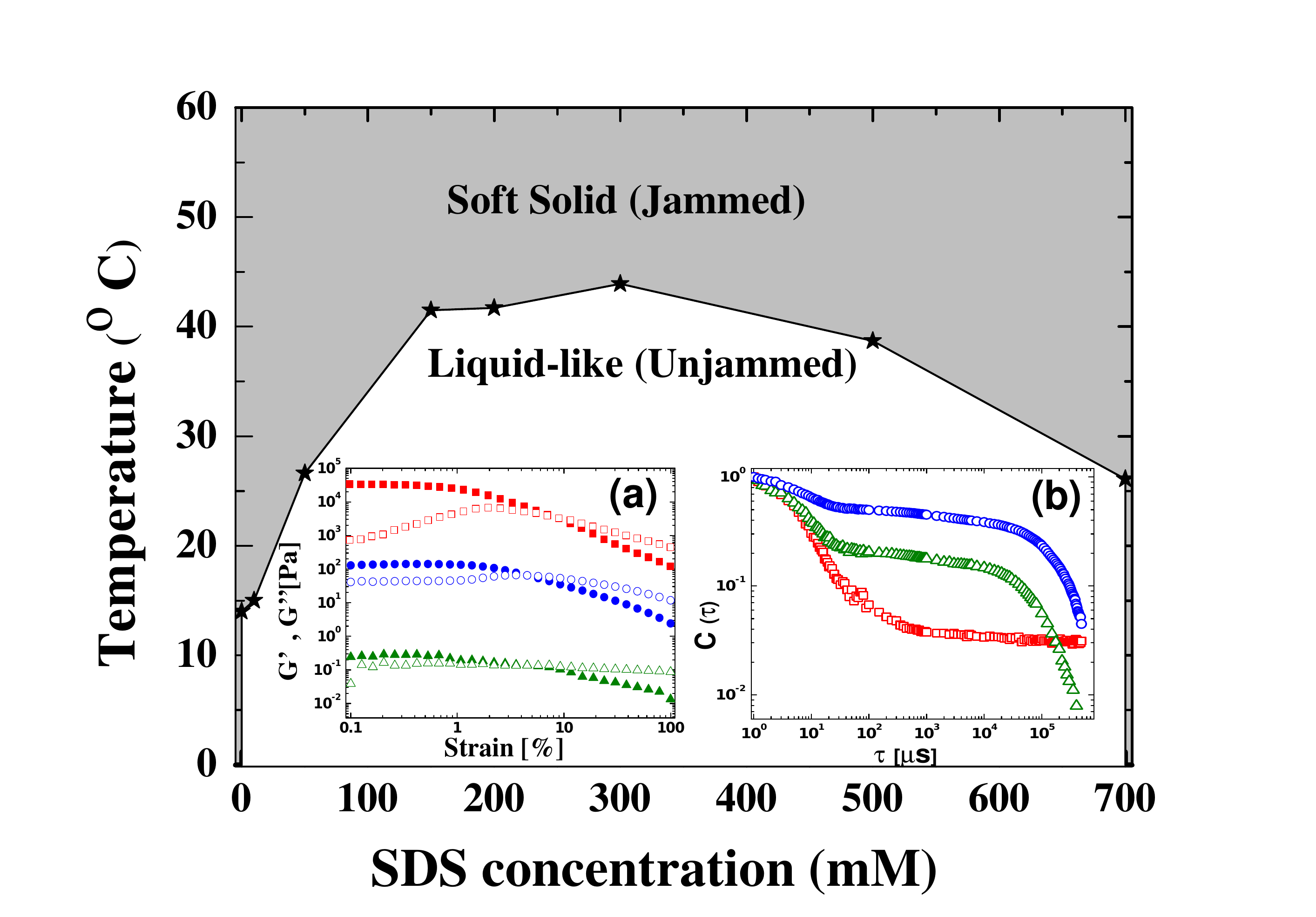}
\caption{Phase diagram in the temperature-SDS concentration plane when F127 concentration is fixed at 0.25 g/cc. The grey region denotes the jammed soft solid-like phase while the white region represents the liquid-like (unjammed) regime. Inset (a) shows the frequency responses at 42$^{\circ}$C for a pure F127 solution (squares) and for F127-SDS mixtures with SDS concentrations 150 mM (circles) and 700 mM (triangles). The solid symbols denote G$^{\prime}$ and the hollow symbols denote G$^{\prime\prime}$. Inset (b) shows the PCS data for the same samples.\\}
\label{FIG 9}
\end{center}
\end{figure}

To summarise the jamming-unjamming behaviour exhibited by the pure F127 solution and F127-SDS mixtures (F127 concentration is fixed at 0.25 g/cc), a phase diagram is constructed in the temperature-SDS concentration plane (Fig. 7). The grey region in this phase diagram indicates the soft solid regime (comprising jammed aggregates), while the white region represents the liquid-like response that is typically seen below the critical micellization temperature. The phase boundary corresponds to the temperature at which the liquid-solid transition occurs in each sample (representative data presented in Fig. 6) and is found to have a non-monotonic dependence on SDS concentration. The jammed region is characterised by different degrees of disorder and metastability and the elasticity of this phase varies over almost two orders of magnitude as SDS concentration is varied. Inset (a) of Fig. 7 shows the frequency response curves measured at 42$^{\circ}$C for the 0.25 /cc F127 sample (squares) and when 150 mM and 700 mM SDS (circles and triangles respectively) are added to this sample. G$^{\prime}$, denoted by solid symbols, and G$^{\prime\prime}$, denoted by hollow symbols, are measured while decreasing the angular frequency logarithmically from 100 rad/s to 0.1 rad/s. The strain amplitude is held fixed at 0.5\% to ensure linear response in all the samples. As expected, the measured moduli decrease sharply with increasing surfactant concentration. For the pure F127 sample (squares), G$^{\prime} >>$ G$^{\prime\prime}$  and is independent of the angular frequency over the entire measurement range, while G$^{\prime\prime}$ is weakly frequency dependent. These are typical signatures of soft glasses \cite{wyss_prl,miyazaki_epl}. With increasing SDS concentration, the signatures of soft glassy behaviour become much less prominent. In the frequency response data for the F127-SDS mixture with an SDS concentration of 700 mM (triangles in inset (a) of Fig. 7), G$^{\prime}$ and G$^{\prime\prime}$ are almost equal, indicating that the sample displays a weak solid-like behaviour over the entire frequency range investigated. Inset (b) of Fig. 7 shows the PCS data for the same samples. It is observed  that after an initial fast decay, the 0.25 g/cc sample (squares) slows down enormously and does not decay within the experimental time window. With the addition of SDS (150 mM SDS and 700 mM SDS, denoted by circles and triangles respectively), the slow decay becomes faster, which is identified with the onset of micellar unjamming. Our frequency response and PCS data, combined with the amplitude sweep measurements displayed in Fig. 4, confirm the gradual disappearance of soft glassy rheology in F127-SDS mixtures on increasing SDS concentration.

\section{Conclusions}

Photon correlation spectroscopy is employed to study the microscopic dynamics of the aggregates that constitute pure F127 solutions and F127-SDS mixtures. This data is related to the mechanical properties of the samples measured using oscillatory rheology. Above a critical temperature, the F127 copolymers in an aqueous solution self-assemble to form micelles with hydrophobic PPO cores and hydrophilic PEO coronas. The PEO chains, which interact $via$ a soft potential, overlap and compress with increasing copolymer concentration. The jamming behaviour that arises from such macromolecular crowding at $c \ge$ 0.10 g/cc increases the solidity of the samples. By combining PCS and rheology data, it is shown that the appearance of soft glassy rheology in the samples is accompanied by a slowing down of their characteristic relaxation time scales. When sufficient quantities of the anionic surfactant SDS is added to a jammed solution of F127 micelles, the pure F127 micelles break up to form anisotropic mixed micelles that are typically much smaller than pure F127 micelles. These anisotropic mixed aggregates, together with the SDS micelles that exist in solution, increase the polydispersity of the sample, thereby increasing its random close packing fraction. This results in an unjamming of the aggregates in F127-SDS mixtures which manifests as a dramatic decrease in the complex moduli of these mixtures and is accompanied by a disappearance of the signature features of soft glassy rheology. Finally, a phase diagram is constructed in the temperature-SDS concentration plane to summarise the jamming-unjamming behaviour of the micelles comprising F127-SDS mixtures. 

\section{Acknowledgements}
The authors thank P. Harsha Mohan for her assistance during the initial stages of these experiments.

\end{document}